# Cyber Security Operations Educational Gamification Application Listing, 2022

## https://bit.ly/3S260GS

**Sherri Weitl-Harms**
*Computer Science, Design and Journalism*
*Creighton University*

Omaha, NE USA
sherriweitlharms@creighton.edu
ORCID 0000-0002-3653-2928

**Adam Spanier**
*Information Systems and Technology*
*University of Nebraska at Omaha*

Omaha, USA
aspanier@unomaha.edu
0000-0003-3523-1119

**John D. Hastings**
*Beacom College of Computer and Cyber Sciences*
*Dakota State University*

Madison, SD 57042, U.S.A.
John.Hastings@dsu.edu
0000-0003-0871-3622

This listing contains a total of 80 gamification applications (GA)s used in cyber security operations (CSO) undergraduate education, from 74 publications, published between 2007 and June 2022. The listing outlines each GA identified and provides a short overview of each. This listing serves as both a comprehensive repository of existing GAs in cybersecurity undergraduate education, and as a starting point for adding new CSO GAs to the list. **Contact the first author to add a CSO GA to the next version of the list.**

The listing was gathered using a methodical, comprehensive literature review using the Systematic Mapping Study (SMS) process to identify implemented and evaluated GAs in undergraduate CSO education (Weitl-Harms et al, 2023a). The SMS was conducted January-July 2022 and focused primarily on four databases: (1) IEEE Xplore, (2) The ACM Digital Library (3) Scopus, and (4) Taylor and Francis. The databases were selected based upon size, popularity, and relevance to the subject and reputation. Each query was limited to abstract only text.

A logical, replicable search string function was implemented to systematically query targeted databases: Category1 + AND + Category2 + AND + Category3. In the search string, one keyword from each of the three categories was chosen and inserted into the search string before committing to the query. A simple combinatorial function cycled through each unique combination to produce the results. The categories were:

- Category 1: Education: education, learn, train, course, student, teach
- Category 2: Gamification: game, gamification, game based learning
- Category 3: UCSO: cybersecurity, cyber security, cyber security operations, computer security

The results were filtered to return only papers written in English, and since gamification was not defined until 2011, papers written before 2005 were excluded. Literature not included in this listing are: (1) evaluations of labs, (2) physical environment studies, (3) competition studies, (4) tutorials, (5) panels, (6) short studies (two or less pages in length), and (7) posters.

To expand the corpus to find relevant related research, citation and reference snowballing was included for the literature that met the filter criteria and related to GAs or Game-based learning in undergraduate CSO education. To stay true to the focus of the study, the data set was refined to only include literature that described an evaluation of the GA or Game-based learning in undergraduate CSO education. Any literature that did not include an evaluation of the GA in CSO education were excluded from the results reported below.

This listing uses the framing organizational constructs applied to CSO educational GAs in Weitl-Harms et al, (2023b): Enhanced Examination (EE), Visualization of Abstract Ideas (VAI), Missions and Quests (MQ), Simulations (Sim), Aspirational Learning (AL), Dynamic Gamification (DG), and Social and Collaborative Engagement (SGE). Seventeen CSO GAs fit in the EE frame due to their optimization in the analysis of learning progress. Four fit the VAI frame. Twenty-seven existing CSO GAs fit within the MQ frame as CSO education lends itself well to these types of experiences. Nine Sim GAs were successfully deployed in CSO education. Seventeen CSO GAs fall within the AL GA frame, many of these manifesting as CTF missions. Three fit the DG frame and three fit the SGE frame.

## CSO Enhanced Examination (EE) Gamification Applications (GAs)

EE1. *Socrative (*Showbie Inc., 2022) is a generalized education GA applied to CSO education in Beltran et al., 2018.

EE2. *Kahoot*! (Kahoot, 2022) is a generalized education GA applied to CSO education in Beltran et al., 2018.

EE3. *Seppo* (Seppo, 2022) is a generalized education GA applied to CSO education in Beltran et al., 2018).

EE4. *OneUP (*Dicheva et al., 2017) generalized gamification framework was also applied to CSO education ( Demmese et al., 2020).

EE5. *UltraLearn* (Raisi et al., 2021) is a platform similar to *OneUp*, that was designed to teach cybersecurity to learners with any background.

EE6. *GamifiedLearn* (Alshammari, 2019) is another similar e-learning system.

EE7. Shahriar et al. (2022) evaluated the *Narrative Integrated Career Exploration (NICE)* platform with a track related to cybersecurity. NICE incorporates certain aspects of gamification including the completion of discreet evolving tasks with attainable rewards.

EE8. *CYPHER* (Younis et al., 2021) is an open-access MOOC-style learning platform, enabling the delivery of interactive learning content covering essential cryptographic algorithms and their application in security protocols. It has the ability to tailor learning content according to students' needs.

EE9. Hajja and Hunt (2020) present an instruction-based web-platform that allows instructors to create programmable text- and media-based interactive elements and personalized mini-lessons, with gamified elements.

EE10. Flushman et al. (2015) outline a collection of alternate reality exercises that explore a number of security topics for first year CSO students. The study found improved student engagement with increased awareness of security as a discipline.

EE11. Dixit et al. (2018) also evaluated various gamified activities for first year CSO students.

EE12. *Cyber Secured* (Kletenik et al., 2021) uses engaging gameplay and challenges to educate students about concepts such as phishing, malware, encryption and passwords. It was evaluated in an e-commerce first year course. They found evidence of increased interest in cybersecurity, and positive attitudes towards the use of this game to teach and assess cybersecurity material.

EE13. *Bodhi* (Chen & Mao, 2012) is an online two-player game in which each player is shown a piece of code snippet and asked to choose whether their partner would think there is a buffer overflow vulnerability at a given position in the code. At the end of one game, the two players can review all the code snippets for which they do not get the points.

EE14. A malware awareness tool, named *MAL-NETS*, was developed and implemented in Omar et al. (2021). From their experience with MALNETS, students who have become aware of malware become cautious of cyber threats.

EE15. *Be-aware (*Mostafa & Faragallah, 2019) is a 2D multiple-choice question based quiz game created to cover social engineering concepts, and teach students how to detect and avoid social engineering attacks.

EE16. *The KMD Puzzle (*Mostafa & Faragallah, 2019) is a 2D image puzzle game that explores content related to key management. The goal of this game is to let students memorize different key management diagrams while they play puzzles and have fun.

EE17. A game-based learning platform designed to enhance cybersecurity education is presented in Khan et al. (2022). The platform includes a virtual lab for students to complete the necessary tools for practice and a web portal where all challenges and learning materials are hosted. The aim is to not only help students learn at their own pace about different cybersecurity challenges, but also give them the opportunity to gain hacking skills with ethics taken into mind in a much safer environment.

## CSO Visualization of Abstract Ideas (VAI) GAs

VAI1. *hACME* is a GA that aims to teach students software security, specifically within web applications, implemented by Nerbraten and Rostad (2009). Each player works through a series of levels where the goal is to discover the vulnerabilities or flaws in a given HTML page in order to unlock access to the next stage. Upon unlocking a new level, all challenges within the level are immediately available. This allows students to pick their own path to the next level. Users can use hints to work through difficult problems, and, as they progress, are awarded points for completing each challenge. These points are then used to rank the player on a leaderboard.

VAI2. Morreale et al. (2019) implement a generalized, gamified pathway for CS students to visualize their progress in any CS major program including CSO. A game board depicts necessary tasks required for each academic year. Badges are also awarded for different pathways. These include the Ready to Succeed badge, the Road to Graduate School badge, and Academic Mage badges.

VAI3. Riposte (Malone et al., 2021) is a framework for measuring skills demonstrated by students within an active learning setting where the primary focus is on practical expertise. The gamified framework is insecure enough to be "hackable", but secure enough not to be abused and is used to expose students to various security concepts.

VAI4. Zhang et al. (2020) created a web-based interactive visualization tool that aims to help students gain a deeper understanding of buffer overflow concepts. It is played as an online game with an analytics dashboard, leaderboards, quizzes, coins and points.

## CSO Mission and Quest (MQ) GAs

MQ1. *CounterMeasures* (Jordan et al., 2011) is a single player game that provides a game-type environment for learning and practicing security skills through a series of guided missions. As a player completes a mission, the score is incremented according to the difficulty and objectives of the mission. For more difficult missions, the player can seek help, but at the cost of subtracting from the score.

MQ2. *BashDungeon* (Corda et al., 2019) is a game designed with adventure inside a dungeon, aimed at reproducing the topology of a Unix file system. Inside the different rooms, the players can learn how to use several Unix commands, from simple file system actions to complex text manipulations, to complete the quests and win the game.

MQ3. *Temple of Treasures* (Weanquoi et al., 2021) is an online 2D educational game that aims to help students learn the basic concepts of Discretionary Access Control and Mandatory Access Control. The game's story is centered around an adventurer who is in search of gold, stuck in a temple, and needing to gain knowledge on targeted concepts to unlock the doors along the escape pathways. It has leaderboards, level controls, and an analytics dashboard.

MQ4. *SherLOCKED* (Jaffray et al., 2021) is a serious game created in the style of a 2D top-down puzzle adventure. The game is used to consolidate students' knowledge of foundational security concepts (e.g. the CIA triad, security threats and attacks and risk management).

MQ5. *Bird's Life* is a 2D game created to help students understand the concepts of phishing (Weanquoi et al., 2018).

MQ6. A "*role-game of the Internet*" game (Catuogno & Santis, 2008) was designed as part of the lab activity of a Network Security course. In this game, instead of fighting against each other, student-teams had to cooperate in order to accomplish a list of business-like tasks over a simulation of the Internet while preserving the security and availability of featured network services.

MQ7. *NITE Team 4* (Alice & Smith, 2022), is a commercially available hacking simulation and strategy game with Alternate Reality game elements, was used to enhance learning in Karagiannis & Magkos (2020). The student evaluation was conducted outside the game environment.

MQ8. Dabrowski et al. (2015) introduce students to real-world security attacks and defense mechanisms through a GA used throughout the course. Each challenge is embedded into a small (typically funny) story line including secret missions, big companies, helicopters, or the image of boring office workers turning into computer security superheroes at night.

MQ9. Schreuders and Butterfield (2016) created and evaluated an online gamified learning environment, called *My XP*, with all assigned learning activities defined in terms of quests with XP rewards.

MQ10. Tioh et al. (2019) created an adventure game for teaching social engineering concepts. The basic premise of the game places players in the shoes of a penetration tester on his first day on the job, whose typical aim is to infiltrate the building of a fictional business and attempt to steal sensitive corporate and/or technical information.

MQ11. *Image, Preserve, Analyze, and Report (IPAR)* (Pan et al., 2017) is a GA that allows students to repeatedly practice forensics tools and reinforce the forensics concepts through detective case studies. Each case is associated with one digital crime scene investigation, and there is a visual representation of the tasks and/or questions to solve each mystery.

MQ12. *Digital Forensics Interactive (DFI)* (Yerby et al., 2014) is a 3D game environment to educate users on digital forensics by giving cases to investigate. The user has to follow the normal digital forensics process to solve the case.

MQ13. *Digital Forensic Game Framework (DFGF)}* (Pan et al., 2015) is a GA that allows students to repeatedly practice forensics tools and reinforce the forensics concepts through detective case studies. Each case is associated with one digital crime scene investigation, and there is a visual representation of the tasks and/or questions to solve each mystery.

MQ14. *Cyberspace Odyssey* (Graham et al., 2020) is a serious game that engages students in a race to successfully perform various cybersecurity tasks in order to collect clues and solve a puzzle in a virtual near-Earth 3D space.

MQ15. The online game *Werewolves of Miller's Hollow* (Ensafi et al., 2012) has been deployed to help students understand information flow. In the game, to avoid being eaten, students must exploit inference channels on a Linux system to discover "werewolves" among a population of "townspeople." Because the werewolves must secretly discuss and vote about who they want to eat at night, they are forced to have some amount of keystroke and network activity in their remote shells at this time. In each instance of the game the werewolves are chosen at random from among the townspeople, creating an interesting dynamic where students must think about information flow from both perspectives and keep adapting their techniques and strategies throughout the semester.

MQ16. *GenCyberCoin* (Ford & Siraj, 2019) is an open-source web platform that provides students with opportunities to earn and spend digital currency, practice bug hunting, and get rewarded for helping peers and completing tasks. This platform introduces students to real-world concepts such as the blockchain, digital currency markets, banks, cyber-security principles, open source intelligence gathering, passwords, bug bounty, and social norms and values.

MQ17. *Security Requirement Education Game (SREG)* (Yasin et al., 2018) is a virtual card game for security requirements education. The game has adaptable maps to give changeability to the game. In one situation, the players are instructed to go and evaluate a particular hospital's organizational and informational settings, obtain vulnerability/weakness and, finally, compromise it by suggesting concrete attack scenarios. The players are working as a team with a common goal to achieve. However, there is competition with other teams. Successful attack scenarios, by analyzing vulnerability and situation for assets, are the winning criteria.

MQ18. *Info-Sec Consultant* (Mostafa & Faragallah, 2019) is a 3D role-playing game similar to CyberProtect designed specifically for CSO undergraduate education. It introduces the logical security techniques to protect computer systems against attacks. The authors also reproduced the traditional board based Snakes and Ladders game in a 2D electronic video game and applied it in the security policy domain.

MQ19. *Quantum Corp* is a GA that uses storytelling to match principal cybersecurity concepts with metaphors for students to solve (Ros et al. 2020).

MQ20. *Anti-Phishing Phil* (Sheng et al., 2007) (and *Anti-Phising Phyllis*) (CMU, 2022) were developed at Carnegie Mellon University to provide user- friendly tools to teach about phishing attacks. In the *Anti-Phishing Phil* game, players have to guide a fish towards different worms that will display a genuine or a phishing link. In the *Anti-Phishing Phyllis* game, players help Phyllis teach her school of fish how to avoid phishing traps in fraudulent emails. Unfortunately, *Anti-Physing Phyllis* was not evaluated for undergraduates.

MQ21. Borrego et al. (2017) created an escape room game activity in an Information and Security course, for students to learn concepts such as information measurement, data compression techniques, cryptography, privacy, authenticity, accessibility and public key and private key infrastructure.

MQ22. DeBello et al. (2022) created *Escape the Classroom* that covered many CSO concepts.

MQ23. Deeb and Hickey (2019) created a 3D Escape Room game for introducing computer security and cryptography.

MQ24. *CySecEscape 2.0* (Loffler et al., 2021) is a virtual escape room addressing the cybersecurity challenges of small and medium-size companies, based on an earlier version of the system created as a physical escape room.

MQ25 Taladriz (2021) created an escape room activity to cover networking concepts.

MQ26. Williams (2021) created a concept map that outlined the relationships of gamification, escape rooms, and learning skills to help future researchers transition content to virtual escape room environments. Their model incorporated the cybersecurity-related skills of social engineering, password security, and binary to create a collaborative virtual experience.

MQ27. The *EDURange* framework (Weiss et al., 2015) is a cloud-based resource for hosting on-demand interactive cybersecurity scenarios. The scenarios they have implemented were designed specifically to nurture the development of analysis skills in students as a complement to both theoretical security concepts and specific software tools. They implemented several exercises in the framework, including exercises for students to learn about mapping a network and understanding network protocols, such as TCP, UDP, ICMP; and exercises for students to learn about intrusion detection and prevention; exercises for students to learn about forensics and reverse engineering; exercises where the players must find data on a target host that is behind a gateway by passively examining network traffic and crafting packets to reveal specific information in a text-based adventure; exercises where the player has to create a set of rules to control traffic in and out of a network; exercises where the defender is given the grammar for a calculator and must implement an interpreter for that grammar and the attacker tries to fuzz the interpreter to produce incorrect results or get it to reject a valid expression; and exercises where students learn to filter large amounts of data to distinguish between normal and anomalous behavior indicative of malware.

## CSO Simulation (Sim) GAs

Sim1. In research carried out by Chothia et al. (2017), a fictional story is used, where students play the part of a new IT security employee at a company and are asked to complete a number of security tasks. In response to completing these tasks, each student receives a flag. The students can send the flags they find to a number of different characters to move the story along in different ways. As the story unfolds they find deceit, corruption and ultimately murder, and their choices lead them to one of three different endings.

Sim2. *Playground* (Nielson, 2016), is a network security simulation and training tool. Students use Playground to create their own network security architecture, almost from the ground up. Upon the completion of a given topology, the students then turn around and figure out all the different ways they might crack it.

Sim3. *PenQuest* (Luh et al., 2020), is a meta model designed to present a complete view on information system attacks and their mitigation while simultaneously providing a tool for both semantic data enrichment and security education. It simulates time-enabled attacker/defender behavior as part of a dynamic, imperfect information multiplayer game that derives significant parts of its ruleset from established information security sources. Attack patterns, vulnerabilities, and mitigating controls are mapped to counterpart strategies and concrete actions. The gamified model considers and defines a wide range of actors, assets, and actions, thereby enabling the assessment of cyber risks while giving technical experts the opportunity to explore specific attack scenarios in the context of an abstracted IT infrastructure.

Sim4. *Cybermatics* is an interactive simulation that allows students to "play" through an authentic scenario (case study) as a member of a professional team (Giboney et al., 2021). The applications saw increased student understanding about certain key aspects of professional cybersecurity work, improved their confidence in being able to successfully apply certain skills associated with cybersecurity, and increased nearly half of the students' interest in pursuing a cybersecurity career.

Sim5. *Simulated Critical Infrastructure Protection Scenarios (SCIPS)* (O'Connor et al., 2021) is an experiential serious game utilized to shape the risk thinking of participants with respect to different cybersecurity scenarios in order to train situational awareness and mental models for incident response. The SCIPS platform is data-driven for cross-extensibility allowing it to be adapted to a whole range of different training requirements.

Sim6. Koch et al. (2012) present a cybersecurity educational application that features a game goal unrelated to IT security. However, during the game session gradually more and more attacks on the underlying infrastructure disturb game play. Such a scenario is very close to the reality of an IT security expert, where establishing security is just a necessary requirement to reach the company's goals.

Sim7. *QuaSim* (Parakh et al., 2017) is designed to educate junior/senior undergraduate and graduate students in quantum cryptographic principles. It poses quantum cryptographic problems developed by domain experts and students are able to interactively find solutions to them. QuaSim facilitates collaborative and competitive project-based student learning of quantum principles.

Sim8. *Space Fighter* (Mostafa & Faragallah, 2019) is a 3D action/adventure game designed to cover phishing attack techniques as well as different types of malware.

Sim9. *Hacking Simulato*r (Mostafa & Faragallah, 2019) is a 2D simulation game designed to simulate network attacks and teach students basic IP/TCP attacks against computer networks.

## CSO Aspirational Learning (AL) GAs

AL1. Alothman et al. (2022) present a *Kuwait cyber range* (*Q8CR*) that runs a cyber-security attack and defense simulation, with red and blue teams that work against each other, and a black team that is responsible for creating and evaluating the scenario cases for both the red and blue teams. This, and other cyber range activities, such as (Vekaria et al., 2021) contain many gamification characteristics.

AL2. *CyberCIEGE* (Navy Postgraduate School, 2021) is a popular security awareness tool that was developed by the Naval Postgraduate School and other collaborators and has been used extensively in cybersecurity education (Schreuders & Butterfield, 2016; Thompson & Irvine, 2011). The game offers realistic virtual world scenarios in which players have to operate and defend a computer network.

AL3. Karagiannis & Magkos (2021) used a Capture The Flag (CTF) framework and challenges through a linear sequence while simultaneously presenting educational context for the students to engage gradually and acquire the appropriate knowledge and skills.

AL4. Vykopal et al., (2020) created a Jeopardy-style CTF game for individual students that consisted of 8 challenges on the topics of substitution ciphers, hashing, symmetric and asymmetric cryptography, RSA, and cryptanalysis.

AL5. Vykopal et al., (2020) created a second Jeopardy-style CTF game for individual students that consisted of 15 challenges on the topics of network traffic analysis, port knocking, access control, buffer overflow, command injection, format string attack, and SQL injection.

AL6. A virtual-machine (VM) based CTF framework was created by Chothia & Novakovic (2015), for CSO students to complete Jeopardy-style CTF challenges. They also focused on technical skills and understanding and were not based on a specific scenario. For all exercises, students were required to submit written answers describing the steps they took to recover flags from the VM, and — where appropriate — a description of what the vulnerabilities were and how they worked, and an explanation of how they could be fixed.

AL7. Incorporating the *CyberChallenge.IT*, a jeopardy-style CTF that is the leading Italian initiative for introducing young talents to the field of cybersecurity, into undergraduate curriculum (ACM-IEEE Joint Task Force on CyberSecurity Curricula., 2017) was discussed and evaluated in Ferraro et al. (2020).

AL8. Vitorino et al. (2022) presented *StarsCTF*, a Capture the Flag experiment designed to assess player types and their levels of engagement. In a paired experiment, an individual Jeopardy format (called Open World) was used, and a new game mode was developed, called DMC (Dynamics, Mechanics and Components). "The Open World's challenges have elements to satisfy players with high scores in the player type achievement (Challenges, Feedback and Points), and DMC ones have elements to satisfy achievement and immersion player types (Narrative, Progression Restrictions, Challenges, Feedback

and Points)."

AL9. Kornegay et al. (2021) evaluated the *MITRE eCTF*, which takes "a systems approach to security, i.e., it considers both the hardware and the software counterparts under consideration for security analysis." The *eCTF* framework also "provides a balanced approach to cyber-attack and defense strategies." They found that the *eCTF* "allowed students to work in teams, develop critical thinking skills, address complex technical issues associated with real-world applications, and motivate continued learning, and increased research productivity after the course ended."

AL10. Facebook's CTF platform has also been used as a learning and assessment tool in CSO education (Beltran et al., 2018; Chicone et al., 2018).

AL11. *GeoCTF* (Yang et al., 2017) is an educational CTF-style tool designed to raise the level of awareness about the dangers of uncontrolled sharing of location data, and to illustrate prominent location protection techniques.

AL12. *SWaT Security Showdown (S3)* (Antonioli et al., 2017) was a gamified CTF event that was specifically targeted at Industrial Control Systems (ICS) security. S3 implemented challenges that include both theoretical and applied ICS security concepts, using simulated and real ICS infrastructures. The competition included international teams of attackers and defenders both from academia and industry.

AL13. A cloud-based election application provides the scenario for a CTF activity designed to teach students about the potential pitfalls and consequences of cloud misconfiguration (Romano et al., 2021). Students pose as malicious actors who seek to compromise an election application running on a cloud environment.

AL14. Another CTF example (Lehrfeld & Guest, 2016) focuses on a radical animal rights group's wishes to free an animal held in zoo captivity. Their goal in this attack is to compromise access to the zoo's website and then delete animals from the zoo's inventory database. The challenge is divided into three phases that mimic an actual penetration testers methodology: a reconnaissance phase, an exploitation phase, and an execution phase.

AL15. *Haaukins* (Haaukins, 2022), is a cyber security training platform, evaluated by Broholm et al. (2022). *Haaukins* is an "immersive, interactive learning platform, which allows students hands-on, practical experience with cyber security and ethical hacking in an online, virtualized environment."

AL16. *HackTheBox* (HackTheBox, 2022), is a cyber security training platform based on a CTF-style competition that is always available. It also was evaluated by Broholm et al. (2022).

AL17. *PicoCTF* (picoCTF, 2022) is a two week competition that everyone can enter to compete, evaluated for CSO education by Broholm et al. (2022).

## CSO Social and Collaborative Engagement (SCE) GAs

SCE1. *PeerSpace* is a network based collaborative learning environment created by Li et al. (2013). PeerSpace utilizes elements like peer review, project repositories, wikis, profiles, friends, blogs and discussions to build relationships and encourage collaboration between students. It also provides a game section which students can use to better understand the coursework.

SCE2. *Classroom Live* is an undergraduate level GA created for CS students, including CSO students. In the development of this software, students and teachers work together to create an application for communicating generalized CS coursework (Freitas & Freitas, 2013).

SCE3. *Code Defenders* (Clegg et al., 2017), is used to teach software testing in a collaborative way. Attackers create mutant versions of the program and defenders write test cases for the program being tested. As players progress through levels of the game, they incrementally learn and practice testing concepts.

### CSO Dynamic Gamification (DG) GAs

DG1. In a study by Svabensky et al. (2018), students participate in a game-development based learning project that sees the individual creation of different penetration testing games. The students report they enjoyed a unique opportunity to deeply understand the topic and practice their soft skills as they presented their results at a faculty open day event. Their peers, who played the created games, rated the quality and educational value of the games as overwhelmingly positive. While the application of this process sees students interacting with unrelated static gamification iterations, the game development pre-phase contains GA elements.

DG2. McGregor et al. (2022) presented the *Citadel Programming Lab* which comprises a GitLab instance for simulated secure programming tasks and a tower defense game. In this game environment, students first play the tutorial level, which exposes them to the purpose of game and gameplay mechanics. This is followed by the students playing the main level, which exposes them to security metaphors, helps them develop motivation to defend their goal and allows them to earn points. Students can then spend points to unlock upgrades, which some upgrade tiers require solving a programming task and reviewing other solutions.

DG3. In a study by Celeda et al. (2020), students participate in a game-development based learning project where paired students create CTF games that are deployed to the Kypoindustry industrial control systems testbed. Then a public hacking day is organized for other students of the university to play the created games. Unfortunately, because the study does not provide an evaluation of the impact of this activity on the students, it is not included in the reported results in Table 1.

### CSO Collaborative Gamification Development (CGD) GAs

GCD0. Currently, *Classroom Live* contains elements of CGD, yet falls within the SCE frame as its primary characteristic attributes placed it within that category (Freitas & Freitas, 2013).

## REFERENCES

Alice & Smith. (2022). Nite team 4. Retrieved March 10, 2022 from https://www.niteteam4.com/
Alothman, B., Alhajraf, A., AlajmiR., Farraj, R. A., Alshareef, N., & Khan, M., (2022). Developing a Cyber Incident Exercises Model to Educate Security Teams. *Electronics, 11*(10), 1575. https://doi.org/10.3390/electronics11101575
Alshammari, M. T. (2019). Design and Learning Effectiveness Evaluation of Gamification in e-Learning Systems. *International Journal of Advanced Computer Science and Applications, 10*(9). https://doi.org/10.14569/ijacsa.2019.0100926
Antonioli, D., Ghaeini, H. R., Adepu, S., Ochoa, M., & Tippenhauer, N. O. (2017). Gamifying ICS security

training and research: Design, implementation, and results of S3. *Proceedings of the 2017 Workshop on Cyber-Physical Systems Security and PrivaCy*, 93–102. https://doi.org/10.1145/3140241.3140253

Beltran, M., Calvo, M., & Gonzalez, S. (2018). Experiences using capture the flag competitions to introduce gamification in undergraduate computer security labs. *2018 International Conference on Computational Science and Computational Intelligence (CSCI)*, 574–579. https://doi.org/10.1109/CSCI46756.2018.00116

Borrego, C., Fernandez, C., Blanes, I., & Robles, S. (2017). Room escape at class: Escape games activities to facilitate the motivation and learning in computer science. *Journal of Technology and Science Education*, *7*(2), 162–171. https://doi.org/10.3926/jotse.247

Broholm, R., Christensen, M. & Sørensen, L. T.. (2022). Exploring Gamification Elements to Enhance User Motivation in a Cyber Security Learning Platform Through Focus Group Interviews, *2022 IEEE European Symposium on Security and Privacy Workshops (EuroS&PW)*, pp. 470-476, doi: 10.1109/EuroSPW55150.2022.00056

Carlisle, M., Chiaramonte, M., & Caswell, D. (2015). Using CTFs for an undergraduate cyber education. *2015 USENIX Summit on Gaming, Games, and Gamification in Security Education (3GSE 15)*, 1–6. https://www.usenix.org/conference/3gse15/summit-program/presentation/carlisle

Carnegie Mellon University (CMU). (2022). *Anti-Phishing Phyllis - Information Security Office - Computing Services - Carnegie Mellon University*. Retrieved March 10, 2022, from https://www.cmu.edu/iso/aware/phyllis/index.html

Catuogno, L., & Santis, A. D. (2008). An internet role-game for the laboratory of network security course. *Proceedings of the 13th Annual Conference on Innovation and Technology in Computer Science Education*, 240–244. https://doi.org/10.1145/1384271.1384336

Celeda, P., Vykopal, J., Svabensky, V., & Slavícek, K. (2020). KYPO4INDUSTRY: A Testbed for Teaching Cybersecurity of Industrial Control Systems. *Proceedings of the 51st ACM Technical Symposium on Computer Science Education*. Association for Computing Machinery, New York, NY, USA, 1026–1032.

Chang, C.-H. S., Kuo, C.-C., Hou, H.-T., & Koe, J. J. Y. (2022). Design and evaluation of a multi-sensory scaffolding gamification science course with mobile technology for learners with total blindness. *Computers in Human Behavior*, *128*, 107085.

Chen, J., & Mao, X. (2012). Bodhi: Detecting buffer overflows with a game. *2012 IEEE Sixth International Conference on Software Security and Reliability Companion*, 168–173. https://doi.org/10.1109/SERE-C.2012.35

Chicone, R., Burton, T., & Huston, J. (2018). Using facebook's open source capture the flag platform as a hands-on learning and assessment tool for cybersecurity education. *International Journal of Conceptual Structures and Smart Applications (IJCSSA)*, *6*(1), 15, https://doi.org/10.4018/IJCSSA.2018010102

Chothia, T., Holdcroft, S., Radu, A.-I., & Thomas, R. J. (2017). Jail, hero or drug lord? turning a cyber security course into an 11 week choose your own adventure story. *2017 USENIX Workshop on Advances in Security Education (ASE 17)*, 1–11. https://www.usenix.org/conference/ase17/workshop- program/presentation/ chothia

Chothia, T., & Novakovic, C. (2015). An offline capture the Flag-Style virtual machine and an assessment of its value for cybersecurity education. *2015 USENIX Summit on Gaming, Games, and Gamification in Security Education (3GSE 15)*, 1–8. https://www.usenix.org/conference/3gse15/summit- program/presentation/ chothia

Clegg, B. S., Rojas, J. M., & Fraser, G. (2017). Teaching software testing concepts using a mutation testing game. *2017 IEEE/ACM 39th International Conference on Software Engineering: Software Engineering Education and Training Track (ICSE-SEET)*, 33–36. https://doi.org/10.1109/ICSE-SEET.2017.1

Corda, F., Onnis, M., Pes, M. Spano, D., & Scateni, R.. (2019). Bashdungeon. *Multimed Tools Appl*, 78, 13731–13746. https://doi.org/10.1007/s11042-019-7230-3

CTFtime Team, C. T. (2021). CTF time. Retrieved on March 14, 2022, from https://ctftime.org/

Dabrowski, A., Kammerstetter, M., Thamm, E., Weippl, E., & Kastner, W. (2015). Leveraging competitive gamification for sustainable fun and profit in security education. *2015 USENIX Summit on Gaming, Games, and Gamification in Security Education (3GSE 15)*, 1–8. https://www.usenix.org/ conference/3gse15/summit-program/presentation/dabrowski

DeBello, J. E., Schmeelk, S., Dragos, D. M., Troja, E., & Truong, L. M. (2022). Teaching effective Cybersecurity through escape the classroom paradigm. *2022 IEEE Global Engineering Education Conference (EDUCON)*. https://doi.org/10.1109/educon52537.2022.9766684

Deeb, F. A., & Hickey, T. J. (2019). Teaching introductory cryptography using a 3d escape-the-room game. *2019 IEEE Frontiers in Education Conference (FIE)*, 1–6. https://doi.org/10.1109/FIE43999.2019.9028549

Demmese, F., Yuan, X., & Dicheva, D. (2020). Evaluating the effectiveness of gamification on students' performance in a cybersecurity course. *Journal of the Colloquium for Information System Security Education*,

*8*(1), 1–12. https://par.nsf.gov/biblio/10290874

Dicheva, D., Irwin, K., & Dichev, C. (2017). Oneup learning: A course gamification platform. In: Dias, J., Santos, P., Veltkamp, R. (eds) *Games and Learning Alliance.* GALA 2017. Lecture Notes in Computer Science(), vol 10653. Springer, Cham. https://doi.org/10.1007/978-3-319-71940-5_14

Dixit R.K., Nirgude M., & Yalagi P.S. (2018). Employ gamification to make "I&CS" more interesting. *Journal of Engineering Education Transformations, 32,* 55-60.

Ensafi, R., Jacobi, M., & Crandall, J. R. (2012). Students who don't understand information flow should be eaten: An experience paper. *CSET*, 1–10.

Flushman, T., Gondree, M., & Peterson, Z. N. J. (2015). This is not a game: Early observations on using alternate reality games for teaching security concepts to First-Year undergraduates. *8th Workshop on Cyber Security Experimentation and Test (CSET 15)*, 1–8, https://www.usenix.org/conference/cset15/workshop-program/presentation/flushman

Ford, V., & Siraj, A. (2019). Gencybercoin: An engaging, customizable, and gamified web platform for cybersecurity summer camps and classrooms. *J. Comput. Sci. Coll.*, *35*(3), 87–96.

Ferraro, G., Lagorio, G., & Ribaudo, M. (2020). CyberChallenge.IT@Unige: Ethical Hacking for Young Talents. *In Adjunct Publication of the 28th ACM Conference on User Modeling, Adaptation and Personalization* (UMAP '20 Adjunct). Association for Computing Machinery, New York, NY, USA, 127–134. https://doi.org/10.1145/3386392.3399311

Freitas, A. A. D., & Freitas, M. M. D. (2013). Classroom live: A software-assisted gamification tool. *Computer Science Education*, *23*(2), 186–206.

Giboney, J., McDonald, J., & Balzotti, J. (2021). Increasing cybersecurity career interest through playable case studies. *TechTrends*, *65*, 496–510. https://doi.org/10.1007/s11528-021-00585-w

Graham, K., Anderson, J., Rife, C., Heitmeyer, B., R. Patel, P., Nykl, S., C. Lin, A., & D. Merkle, L. (2020). Cyberspace odyssey: A competitive team-oriented serious game in computer networking. *IEEE Transactions on Learning Technologies*, *13*(3), 502–515. https://doi.org/10.1109/TLT.2020.3008607

HackTheBox, (2022). https://hackthebox.com/

Hajja, A., & Hunt, A. J. (2020). A novel e-learning platform for building and publishing student-driven personalized lessons. *2020 IEEE Frontiers in Education Conference (FIE)*, 1–8. https://doi.org/10.1109/FIE44824.2020.9274034

Haaukins, (2022). https://general.haaukins.com/

Hellerstedt, A., & Mozelius, P. (2019). Game-based learning: A long history. *Irish Conference on Game-based Learning 2019, Cork, Ireland, June 26-28, 2019*, *1*, 1–4.

Jaffray, A., Finn, C., & Nurse, J. (2021). Sherlocked: A detective-themed serious game for cyber security education. *Human Aspects of Information Security and Assurance, HAISA 2021*, *613*, 34–45. https://doi.org/10.1007/978-3-030-81111-2_4

Jordan, C., Knapp, M., Mitchell, D., Claypool, M., & Fisler, K. (2011). Countermeasures: A game for teaching computer security. *2011 10th Annual Workshop on Network and Systems Support for Games*, 1–6. https://doi.org/10.1109/NetGames.2011.6080983

Kahoot, (2022). Kahoot!. https://kahoot.com/

Karagiannis, S., & Magkos, E. (2020). Adapting CTF challenges into virtual cybersecurity learning environments. *Information & Computer Security, 29*(1), 105–132. https://doi.org/10.1108/ics-04-2019-0050

Karagiannis, S., & Magkos, E. (2021). Engaging students in basic cybersecurity concepts using digital game-based learning: Computer games as virtual learning environments. Springer. https://doi.org/10.1007/978-3-030-41196-14

Khan, M. A., Merabet, A., Alkaabi, S., & Sayed, H. E. (2022). Game-based learning platform to enhance cybersecurity education. *Education and Information Technologies, 27*(4), 5153–5177. https://doi.org/10.1007/s10639-021-10807-6

Kletenik, D., Butbul, A., Chan, D., Kwok, D., &LaSpina, M. (2021). Game on: teaching cybersecurity to novices through the use of a serious game. *J. Comput. Sci. Coll. 36, 8* (April 2021), 11–21.

Koch, S., Schneider, J., & Nordholz, J. (2012). Disturbed playing: Another kind of educational security games. *5th Workshop on Cyber Security Experimentation and Test (CSET 12)*, 1–9. https://www.usenix.org/conference/cset12/workshop-program/presentation/Koch

Kornegay, M. A., Arafin, M. T., & Kornegay, K., (2021). Engaging Underrepresented Students in Cybersecurity using Capture-the-Flag(CTF) Competitions (Experience). *ASEE Annual Conference and Exposition,* Conference Proceedings.

Lehrfeld, M., & Guest, P. (2016). Building an ethical hacking site for learning and student engagement.

*SoutheastCon 2016*, 1–6. https://doi.org/10.1109/SECON.2016.7506746

Li, C., Dong, Z., Untch, R. H., & Chasteen, M. (2013). Engaging computer science students through gamification in an online social network based collaborative learning environment. *International Journal of Information and Education Technology*, *3*(1), 72.

Li, C., & Kulkarni, R. (2016). Survey of cybersecurity education through gamification. *ASEE Annual Conference and Exposition,* Conference Proceedings. https://www.scopus.com/inward/record.uri?eid=2-s2.0-84983372372&partnerID=40&md5=ffa680de2d5bf647dec56cf13fcdb2cf

Loffler, E., Schneider, B., Zanwar, T., & Asprion, P. M. (2021). Cysecescape 2.0—a virtual escape room to raise cybersecurity awareness. *International Journal of Serious Games*, *8*(1), 59–70. https://doi.org/10.17083/ijsg.v8i1.413

Luh, R., Temper, M., Tjoa, S., Schrittwieser, S., & Janicke, H. (2020). Penquest: A gamified attacker/defender meta model for cyber security assessment and education. *J Comput Virol Hack Tech*, *16*, 19–61. https://doi.org/10.1007/s11416-019-00342-x

Malone, M., Wang, Y., James, K., Anderegg, M., Werner, J., & Monrose, F. (2021). To gamify or not? on leaderboard effects, student engagement and learning outcomes in a cybersecurity intervention. *Proceedings of the 52nd ACM Technical Symposium on Computer Science Education*, 1135–1141. https://doi.org/10.1145/3408877.3432544

McGregor, L., Chan, S. C., Wlodarczyk, S., & Maarek, M. (2022). Aligning a Serious Game, Secure Programming and CyBOK-Linked Learning Outcomes. *2022 IEEE European Symposium on Security and Privacy Workshops (EuroS&PW)*. https://doi.org/10.1109/eurospw55150.2022.00058

Morreale, P., Diplan, N., & York, D. (2019). A gamification pathway for computer science student success. *ITiCSE '19: Proceedings of the 2019 ACM Conference on Innovation and Technology in Computer Science Education*, 317. https://doi.org/https://doi.org/10.1145/3304221.3325577

Mostafa, M., & Faragallah, O. S. (2019). Development of serious games for teaching information security courses. *IEEE Access*, *7*, 169293–169305. https://doi.org/10.1109/ACCESS.2019.2955639

Navy Postgraduate School. (2021). Incorporating cyberciege into an introductory cyber security course. *Naval Postgraduate School*. Retrieved on March 14, 2022, from https://nps.edu/web/c3o/syllabus

Nerbraten, Y., & Rostad, L. (2009). Hacmegame: A tool for teaching software security. *2009 International Conference on Availability, Reliability and Security*, 811–816. https://doi.org/10.1109/ARES.2009.135

Nielson, S. J. (2016). Playground: Preparing students for the cyber battleground. *Computer Science Education*, *26*(4), 255–276. https://doi.org/10.1080/08993408.2016.1271526

O'Connor, S., Hasshu, S., Bielby, J., Colreavy-Donnelly, S., Kuhn, S., Caraffini, F., & Smith, R. (2021). Scips: A serious game using a guidance mechanic to scaffold effective training for cyber security. *Information Sciences*, *580*, 524–540. https://doi.org/https://doi.org/10.1016/j.ins.2021.08.098

Order of the Overflow. (2021). Defcon capture the flag qualifier. *Order of the Overflow.* Retrieved on March 14, 2022, from https://oooverflow.io/dc-ctf-2021-quals/

Omar, N. S., Foozy1, C., Hamid, I., Hafit, H., Arbain, A., & Shamala, P. (2021). Malware awareness tool for internet safety using gamification techniques. *Journal of Physics: Conference Series*, *1874*, 1–9.

Pan, Y., Mishra, S., & Schwartz, D. (2017). Gamifying Course Modules for Entry Level Students. *In Proceedings of the 2017 ACM SIGCSE Technical Symposium on Computer Science Education* (SIGCSE '17). Association for Computing Machinery, New York, NY, USA, 435–440. https://doi.org/10.1145/3017680.3017709

Pan, Y., Schwartz, D.& Mishra, S., (2015). Gamified digital forensics course modules for undergraduates. *In 2015 IEEE Integrated STEM Education Conference*. IEEE, 100-105. https://doi.org/10.1109/ISECon.2015.7119899

Parakh, A., Subramaniam, M., & Ostler, E. (2017). Quasim: A virtual quantum cryptography educator. *2017 IEEE International Conference on Electro Information Technology (EIT)*, 600–605. https://doi.org/10.1109/EIT.2017.8053434

picoCTF, (2022). https://picoctf.org/

Raisi, S., Ghasemshirazi, S., & Shirvani, G. (2021). UltraLearn: Next-Generation CyberSecurity Learning Platform. *2021 12th International Conference on Information and Knowledge Technology (IKT)*. https://doi.org/10.1109/ikt54664.2021.9685940

Romano, Z., Windsor, J., VanDerPol, M., & Coffman, J. (2021). Election security in the cloud: A ctf activity to teach cloud and web security. *2021 IEEE Frontiers in Education Conference (FIE)*, 1–5. https://doi.org/10.1109/FIE49875.2021.9637368

Ros, S., Gonzalez, S., Robles, A., Tobarra, L., Caminero, A., & Cano, J. (2020). Analyzing Students' Self-Perception of Success and Learning Effectiveness Using Gamification in an Online Cybersecurity Course. *IEEE Access, 8*, 97718–97728. https://doi.org/10.1109/access.2020.2996361

Shahriar, S., Ramesh, J., Towheed, M., Ameen, T., Sagahyroon, A., & Al-Ali, A. R. (2022). Narrative Integrated Career Exploration Platform. *Frontiers in Education*, 7. https://doi.org/10.3389/feduc.2022.798950

Schreuders, Z. C., & Butterfield, E. (2016). Gamification for teaching and learning computer security in higher education. *2016 USENIX Workshop on Advances in Security Education (ASE 16)*, 1–8. https://www.usenix.org/conference/ase16/workshop-program/presentation/schreuders

Seppo. (2022). Seppo - Spark for Learning. Retrieved March 10, 2022 from http://seppo.io/

Shellphish.net (2021). iCTF: the International Capture The Flag Competition, Retrieved March 10, 2022, from https://shellphish.net/ictf/

Sheng, S., Magnien, B., Kumaraguru, P., Acquisti, A., Cranor, L. F., Hong, J., & Nunge, E. (2007). Anti-phishing phil: The design and evaluation of a game that teaches people not to fall for phish. *Proceedings of the 3rd Symposium on Usable Privacy and Security*, 88–99. https://doi.org/10.1145/1280680.1280692

Showbie Inc. (2022). Sacrative . Retrieved March 10, 2022 from https://www.socrative.com

Svabensky, V., Vykopal, J. and Celeda, P. (2020). What are cybersecurity education papers about? a systematic literature review of sigcse and iticse conferences. *In Proceedings of the 51st ACM Technical Symposium on Computer Science Education*, 2-8.

Svabensky, V., Vykopal, J., Cermak, M., & Lastovicka, M. (2018). Enhancing cybersecurity skills by creating serious games. *Proceedings of the 23rd Annual ACM Conference on Innovation and Technology in Computer Science Education*, 194–199. https://doi.org/10.1145/3197091.3197123

Taladriz, C. C. (2021). Flipped mastery and gamification to teach Computer networks in a Cybersecurity Engineering Degree during COVID-19. *2021 IEEE Global Engineering Education Conference (EDUCON)*. https://doi.org/10.1109/educon46332.2021.9453885

Tangent, D. (2021). Defcon capture the flag. Retrieved on March 14, 2022, from https://defcon.org/html/links/dc-ctf.html

Thompson, M., & Irvine, C. (2011). Active learning with the cyberciege video game. *Proceedings of the 4th Conference on Cyber Security Experimentation and Test*, 10.

Tioh, J.-N., Mina, M., & Jacobson, D. W. (2019). Cyber security social engineers an extensible teaching tool for social engineering education and awareness. *2019 IEEE Frontiers in Education Conference (FIE)*, 1–5. https://doi.org/10.1109/FIE43999.2019.9028369

US Department of Defense (DOD) Cyber Exchange. (2021). Cyber protect. Retrieved March 14, 2022, from https:/public.cyber.mil/training/cyber-protect/

Vekaria, K. B., Calyam, P., Wang, S., Payyavula, R., Rockey, M., & Ahmed, N. (2021). Cyber Range for Research-Inspired Learning of "Attack Defense by Pretense" Principle and Practice. *IEEE Transactions on Learning Technologies, 14*(3), 322–337. https://doi.org/10.1109/tlt.2021.3091904

Vitorino, D., Bittencourt, I. I., & Chalco, G. (2021). StarsCTF: A Capture the Flag Experiment to Hack Player Types and Flow Experience. *Smart Innovation, Systems and Technologies*, 467–477. https://doi.org/10.1007/978-981-16-4884-7_39

Vykopal, J., Svabensky, V., & Chang, E. C. (2020). Benefits and pitfalls of using capture the flag games in university courses. *Proceedings of the 51st ACM technical symposium on computer science education* (pp. 752–758). Association for Computing Machinery. https://doi.org/10.1145/3328778.3366893

Weanquoi, P., Johnson, J., & Zhang, J. (2018). Using a game to improve phishing awareness. *Journal of Cybersecurity Education, Research and Practice*, *2018*(2), 1–14. https://digitalcommons.kennesaw.edu/jcerp/vol2018/iss2/2

Weanquoi, P., Zhang, J., Yuan, X., Xu, J., & Jones, E. J. (2021). Learn access control concepts in a game. *2021 IEEE Frontiers in Education Conference (FIE)*, 1–6. https://doi.org/10.1109/FIE49875.2021.9637228

Weiss, R. S., Boesen, S., Sullivan, J. F., Locasto, M. E., Mache, J., & Nilsen, E. (2015). Teaching cybersecurity analysis skills in the cloud. *Proceedings of the 46th ACM Technical Symposium on Computer Science Education*, 332–337. https://doi.org/10.1145/2676723.2677290

Weitl-Harms, S.K., Spanier, A.M., Hastings, J.D., & Rokusek, M. (2023a). A systematic mapping study on gamification applications for undergraduate cybersecurity education, *Journal of Cybersecurity Education, Research and Practice (JCERP)*, Vol. 2023, No. 1, Article 9, July 2023. https://doi.org/10.32727/8.2023.12

Weitl-Harms, S.K., Spanier, A.M., Hastings, J.D., & Rokusek, M. (2023b). Framing Gamification in Undergraduate Cybersecurity Education, *Journal of The Colloquium for Information Systems Security Education* (JCISSE), vol. 10, no. 1, March 2023. https://doi.org/10.53735/cisse.v10i1.161

Williams, T., & O., E.-G. (2021). Design of a virtual cybersecurity escape room. *National Cyber Summit (NCS) Research Track 2021*, *310*, 1–14.

Yang, J., Niculaescu, O. G., & Ghinita, G. (2017). A game-oriented educational tool for location privacy topics.

*Proceedings of the 25th ACM SIGSPATIAL International Conference on Advances in Geographic Information Systems*. https://doi.org/10.1145/3139958.3140016

Yasin, A., Liu, L., Li, T., Wang, J., & Zowghi, D. (2018). Design and preliminary evaluation of a cyber security requirements education game (sreg). *Information and Software Technology*, *95*, 179–200.

Yerby, J., Hollifield, S., Kwak, M., & Floyd, K. (2014). Development of serious games for teaching digital forensics. *Issues in Information Systems*, *15*(2), 335–343.

Younis Y. A., Kifayat, K., Shi, Q., Matthews, E., Griffiths, G., & Lambertse, R. (2020). Teaching Cryptography Using CYPHER (InteraCtive CrYPtograpHic Protocol TEaching and LeaRning). In Proceedings of the 6th International Conference on Engineering & MIS 2020 (ICEMIS'20). Association for Computing Machinery, New York, NY, USA, Article 12, 1–7. https://doi.org/10.1145/3410352.3410742

Zhang, J., Yuan, X., Johnson, J., Xu, J., & Vanamala, M. (2020). Developing and assessing a web-based interactive visualization tool to teach buffer overflow concepts. *2020 IEEE Frontiers in Education Conference (FIE)*, 1–7. https://doi.org/10.1109/FIE44824.2020.9274239